%% file: main.tex
\documentclass[sigconf, nonacm]{acmart}
\usepackage{amsmath,amsthm}
\usepackage{booktabs}
\usepackage{tabularx}
\usepackage{microtype}
\usepackage{xcolor}
\usepackage{hyperref}
\usepackage{enumitem}
\usepackage{tikz}
\usepackage{balance}
\usetikzlibrary{arrows.meta,positioning,calc,shapes.geometric,fit,backgrounds}

\newcommand{\schele}[1]{\textsf{#1}}
\input{preamble}

\newcommand\vldbyear{2026}
\newcommand\vldbworkshop{15th International Workshop on Quality in Databases (QDB’26)}
\newcommand\vldbauthors{\authors}
\newcommand\vldbtitle{\shorttitle} 
\newcommand\vldbavailabilityurl{}
\newcommand\vldbpagestyle{plain} 

\begin{document}
\title{Towards Inference-Aware Privacy Guidance for Data Preparation}

\author{Vishal Chakraborty}
\affiliation{%
  \institution{University of California, Irvine}
  \country{USA}
}
\email{vchakrab@uci.edu}

\author{Felix Naumann}
\affiliation{%
  \institution{Hasso Plattner Institute, University of Potsdam}
  \country{Germany}
}
\email{felix.naumann@hpi.de}

\renewcommand{\shortauthors}{Chakraborty et al.}

\begin{abstract}
Data preparation often begins with sensitive data and produces a releasable artifact for analysis, sharing, or model training. Existing
workflows are primarily guided by utility: a curator drops attributes, coarsens values, filters populations, and suppresses tuples until the resulting dataset appears useful and safe. Privacy, when considered, is usually evaluated only on the final release. 

We propose privacy-aware data preparation as an interactive guidance problem. We model a preparation plan as a sequence of deterministic curation operators and ask how each step changes the evidence available to an observer with prior knowledge about a target. Our semantics is based on compatibility sets, which
capture the source tuples still plausible for the target after a released representation is observed. This view separates operators that remove evidence from those that remove ambiguity, explains why privacy effects can be non-monotone, and supports prefix-level feedback under a disclosure budget.
The result is an inference-aware foundation for guiding curators throughout data preparation, rather than judging privacy only after the final artifact is produced. We conclude by identifying the key challenges in building interactive, inference-aware data preparation systems.
\end{abstract}

\maketitle

\pagestyle{\vldbpagestyle}
\begingroup\small\noindent\raggedright\textbf{VLDB Workshop Reference Format:}\\
\vldbauthors. \vldbtitle. VLDB \vldbyear\ Workshop: \vldbworkshop.\\ 
\endgroup
\begingroup
\renewcommand\thefootnote{}\footnote{\noindent
This work is licensed under the Creative Commons BY-NC-ND 4.0 International License. Visit \url{https://creativecommons.org/licenses/by-nc-nd/4.0/} to view a copy of this license. For any use beyond those covered by this license, obtain permission by emailing \href{mailto:info@vldb.org}{info@vldb.org}. Copyright is held by the owner/author(s). Publication rights licensed to the VLDB Endowment. \\
\raggedright Proceedings of the VLDB Endowment. 
ISSN 2150-8097. \\
}\addtocounter{footnote}{-1}\endgroup

\ifdefempty{\vldbavailabilityurl}{}{
\vspace{.3cm}
\begingroup\small\noindent\raggedright\textbf{VLDB Workshop Artifact Availability:}\\
The source code, data, and/or other artifacts have been made available at \url{\vldbavailabilityurl}.
\endgroup
}
\input{content/introduction}

\input{content/inference}

\input{content/leakagesurface}

\input{content/taxonomy}

\balance
\bibliographystyle{ACM-Reference-Format}
\bibliography{sample-base}

\end{document}

%% file: preamble.tex
\usepackage[table]{xcolor}  
\usepackage{lipsum}
\usepackage{blindtext}
\usepackage{setspace}
\usepackage{verbatim}
\usepackage{comment}
\usepackage{tcolorbox}

\usepackage{amsmath}           
\usepackage{amsfonts}
\usepackage{amsthm}

\usepackage{amsmath, amssymb}         
\usepackage{mathtools}
\usepackage{nicefrac}
\usepackage{stmaryrd}
\usepackage{mathrsfs}
\usepackage{bbm}
\usepackage{etoolbox}

\usepackage{xspace}

\usepackage{array}
\usepackage{booktabs}
\usepackage{multirow}
\renewcommand{\arraystretch}{0.8}

\newcolumntype{L}[1]{>{\raggedright\arraybackslash}p{#1}}

\usepackage{threeparttable}
\usepackage{colortbl} 

\usepackage{microtype}

\usepackage{enumerate}
\usepackage{enumitem}

\usepackage{graphicx}    
\usepackage{float}
\usepackage{caption}
\usepackage{subcaption}
\usepackage{caption}     
\usepackage{graphicx}    

\usepackage{url}

\usepackage{algorithm}
\usepackage{algpseudocodex}

\usepackage{siunitx}
\definecolor{darkgreen}{rgb}{0.0, 0.5, 0.0}

\usepackage{tikz}
\usetikzlibrary{arrows.meta,calc,decorations.pathreplacing} 
\usetikzlibrary{decorations.text,calc,arrows.meta,backgrounds,shapes,arrows,patterns,shadows}
\usepackage{pgfplots}
\pgfplotsset{compat=1.18}
\usetikzlibrary{calc,patterns,shadows,backgrounds,positioning}

\usetikzlibrary{decorations.text,calc,arrows.meta,backgrounds,shapes,arrows,patterns,shadows,positioning,fit,matrix}
\usepgfplotslibrary{groupplots}

\usepackage{pifont}
\usepackage{csquotes}
\usepackage{eurosym}
\usepackage{balance}
\usepackage{qtree}
\usepackage{chngcntr}

\usepackage{array}
\usepackage{booktabs}

\definecolor{softgray}{RGB}{238,240,243}
\definecolor{softblue}{RGB}{232,240,250}
\definecolor{softgreen}{RGB}{233,245,233}
\definecolor{softred}{RGB}{249,233,233}
\definecolor{softyellow}{RGB}{252,246,221}

%% file: content/introduction.tex
\section{Introduction}
\label{sec:intro}

Data preparation/curation has become a fundamental database workflow: transforming a
"raw" dataset, often containing private or sensitive information, into a
releasable artifact for analysis, sharing, model training, or public use~\cite{haipipe2023, diffprep2023, crisan2023}.
This transformation is rarely a single operation. A curator drops attributes,
coarsens values, selects populations, suppresses tuples, and repeats these
choices until the release appears useful and safe. This operator view is
central to interactive data preparation systems such as Potter's Wheel and
Wrangler~\cite{raman2001potterswheel,kandel2011wrangler}, to
data-transformation-by-example systems~\cite{jin2017foofah}, and to recent
pipeline search and preparation frameworks~\cite{yang2021autopipeline,
haipipe2023,diffprep2023}. In these settings, curation choices are primarily
guided by utility: which attributes are needed, which values should be
generalized, and which records should remain in scope. Privacy, when
considered, is usually evaluated on the resulting release rather than on the
sequence of operators that produced it.

The same operators that
shape utility also determine what evidence remains in the released artifact and
how that evidence changes across curation steps. Dropping a sensitive column
may not remove evidence for the sensitive fact, because retained attributes may
still make the fact predictable. This concern is central to statistical
disclosure control and anonymization, where releases may remain identifying or
revealing even after explicit identifiers are removed~\cite{
samarati2001protecting,sweeney2002kanonymity}. Refinements such as
$\ell$-diversity and $t$-closeness address the fact that
indistinguishability alone may still leave sensitive values predictable within
an equivalence class~\cite{machanavajjhala2007ldiversity,li2007tcloseness}.
More general privacy models and de-anonymization attacks show that auxiliary
information and correlations can make a release revealing even when the
sensitive attribute is absent~\cite{kifer2014pufferfish,narayanan2008}.
Conversely, deleting a tuple may make a release more revealing by removing the
record that made the target ambiguous. Thus privacy is not only a constraint to
check after a useful release is found. A utility-driven search may miss another
feasible plan that is both more useful and less revealing, because disclosure
depends on operator-induced comparison classes, correlations, and ambiguity.

\input{figs/exampleoverview}

\begin{example}[Running example]
\label{ex:running}
Consider the curation steps in Fig.~\ref{fig:running-combined}(a)--(d). A curator
preparing the data for a treatment-pattern study begins with the source relation $R$ in Fig.~\ref{fig:rel-src}, first drops \schele{Diag},
bucketizes \schele{Age}, and, finally, shortens \schele{ZIP}, producing $R_1$~(Fig.~\ref{fig:rel-1}). The
release appears privacy-conscious because the diagnosis column is absent. Yet
an observer who knows that the target is a woman in her thirties in ZIP prefix
926 sees that the matching released records contain two tuples with
\schele{Med}=\schele{ART} and one with \schele{Med}=\schele{None}. If
\schele{ART} is strongly associated with \schele{HIV}, then $R_1$ still
supports the sensitive inference.

Now consider two single-operator refinements. In $R_2$~(Fig.~\ref{fig:rel-2}), the curator also drops
\schele{Med}. This removes the visible indicator and restores the target to
the larger prior group. In $R_3$~(Fig.~\ref{fig:rel-3}), the curator instead keeps \schele{Med} but
deletes $t_3$, the only matching tuple with \schele{Med}=\schele{None}. This
makes the release smaller, but not necessarily safer: it removes ambiguity and
leaves only ART records in the target's visible profile. Thus two ordinary
curation operators that look comparable from a data preparation perspective
can have different privacy effects.
\end{example}

Example~\ref{ex:running} illustrates the central difficulty. Privacy under
curation is not determined only by which sensitive attributes are removed from
the final release. It depends on how each operator changes the observer's
evidence. Projection may remove evidence. Generalization may weaken evidence.
Selection and tuple deletion may either reduce or increase disclosure,
depending on which tuples they remove and how those tuples affected ambiguity.
A final-only privacy check cannot explain which operator caused the change, in
what direction, or by how much.

We adopt an \emph{operator-level view} of privacy-aware data curation. Rather than
asking only whether the final release satisfies a privacy test, we ask how each
curation step changes the observer's evidence about protected secrets. We model
a curation plan as a sequence of deterministic operators over a source relation.
The adversary observes the released data and has prior knowledge about a target,
such as a visible demographic profile. This is close in spirit to privacy
definitions that reason about belief change under side information~\cite{
kifer2011,kifer2014pufferfish,dwork2006calibrating,dwork2010boosting}, but our
mechanisms are ordinary deterministic curation operators rather than randomized
query mechanisms.

Our semantics uses \emph{compatibility sets:} for a target tuple, the
compatibility set contains the source tuples that remain plausible after the
adversary combines prior knowledge with the released representation. While alternative privacy semantics might be suitable, the compatibility sets
expose the operator-level phenomenon. Column and value operators change how
source tuples appear in the release. Row operators change which tuples remain
present. These effects are not interchangeable. In Example~\ref{ex:running},
dropping \schele{Med} expands the target's compatible set and weakens the
evidence for \schele{HIV}; deleting $t_3$ removes contrary evidence and
strengthens, or at least preserves, the inference. The usual vocabulary of
``removing information'' is therefore too coarse: what matters is whether an
operator removes evidence, removes ambiguity, or changes the representation
through which the target is compared to the source.

Existing work does not directly address this setting. Data preparation systems
formalize, synthesize, and explain transformation sequences, but do not attach
privacy semantics to individual preparation operators~\cite{
raman2001potterswheel,kandel2011wrangler,jin2017foofah,yang2021autopipeline,
shraga2023explain,haipipe2023,diffprep2023}. Statistical disclosure control
and Bayesian privacy give semantics to released tables, but typically evaluate
the release as a single object rather than as the result of a step-indexed
construction process~\cite{samarati2001protecting,sweeney2002kanonymity,
machanavajjhala2007ldiversity,li2007tcloseness,kifer2014pufferfish}.
Differential privacy provides powerful composition and accounting principles,
including systems for private data analysis~\cite{dwork2006calibrating,
dwork2010boosting,dwork2016concentrated,bun2016concentrated,mcsherry2009pinq,
rogers2016odometers,zhang2018ektelo}. These frameworks, however, account for
randomized mechanisms, whereas many curation pipelines are deterministic.
PrivateClean brings differential privacy to data cleaning, but still treats
privacy through randomized DP mechanisms rather than through the
operator-specific evidential effects of deterministic curation steps~\cite{
krishnan2016privateclean}. What is missing is a framework that treats
deterministic curation operators themselves as privacy-relevant
transformations. Our work is also related to inference-aware deletion, where the goal is to
remove a value while limiting what remains inferable from dependencies in the
retained database~\cite{chakraborty2025deletion,chakraborty2026deletion}. 
In deletion, the request specifies a target value to remove. 
In curation, the analyst constructs a release through heterogeneous operators, and
the privacy question is how those operators alter the evidence available to an
observer during and after plan construction.

We study privacy-aware curation as an \emph{interactive guidance} problem. Given a
source relation, deterministic curation operators, protected secrets, and an
adversary with prior knowledge about a target, how can a curator construct a
useful release while keeping evidential disclosure below a budget? The challenge
is that disclosure is a property of plan prefixes, not only endpoints. A
release that appears safe at the end may be reached through revealing
intermediate states. A plan that was safe when committed may become unsafe when
the source data evolves. And a curator choosing one operator at a time needs
feedback before committing to an operator, not only after the release is
complete.

This paper makes three contributions.
\begin{itemize}
\item We define a curation and privacy framework in which formulate privacy-aware data curation as an operator-level inference
problem~
(Sec.~\ref{sec:model}).

\item We characterize the disclosure effects of common curation operators and
show how they support local guidance under a disclosure budget~(Sec.~\ref{sec:operators-guidance})

\item We outline research questions for interactive
privacy-aware curation, including per-operator semantics, prefix-level
accounting, budgeted guidance, re-execution under drift, and richer disclosure
models~(Sec.~\ref{sec:challenges}).
\end{itemize}

%% file: figs/exampleoverview.tex
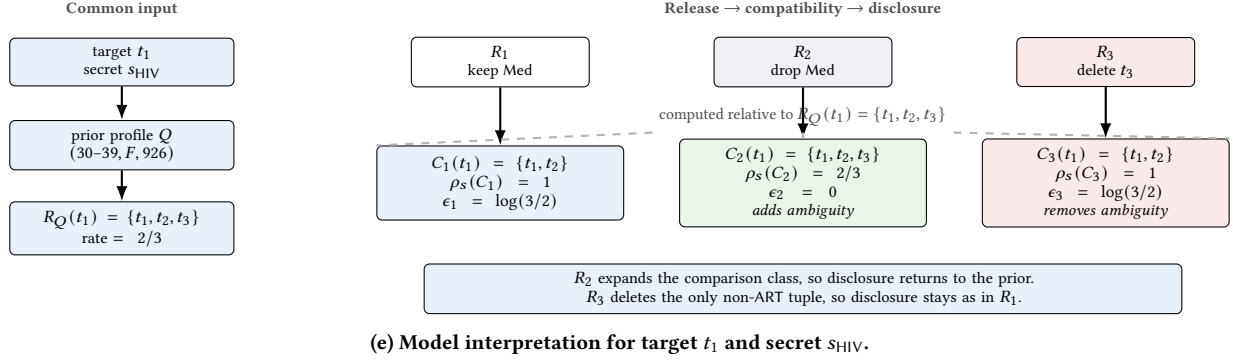
\begin{figure*}[t]
\centering
\footnotesize

\begin{minipage}[t]{\linewidth}
\centering
\setlength{\tabcolsep}{2.5pt}
\renewcommand{\arraystretch}{1.02}

\begin{subfigure}[t]{0.31\linewidth}
\centering
\begin{tabular}{@{}c|c|c|c|c|c@{}}
\schele{pid}&\schele{Age}&\schele{Sex}&\schele{ZIP}&\schele{Med}&\schele{Diagnosis}\\\hline
$t_1$&34&F&92612&\schele{ART}&\schele{HIV}\\
$t_2$&36&F&92617&\schele{ART}&\schele{HIV}\\
$t_3$&35&F&92612&\schele{None}&\schele{Flu}\\
$t_4$&61&M&90024&\schele{Ins.}&\schele{Diab.}
\end{tabular}
\caption{Source $R$.}
\label{fig:rel-src}
\end{subfigure}
\hfill
\begin{subfigure}[t]{0.23\linewidth}
\centering
\begin{tabular}{@{}c|c|c|c@{}}
\schele{Bkt}&\schele{Sex}&\schele{ZIP3}&\schele{Med}\\\hline
30--39&F&926&\schele{ART}\\
30--39&F&926&\schele{ART}\\
30--39&F&926&\schele{None}\\
60--69&M&900&\schele{Ins.}
\end{tabular}
\caption{$R_1$: drop \schele{Diag}.}
\label{fig:rel-1}
\end{subfigure}
\hfill
\begin{subfigure}[t]{0.17\linewidth}
\centering
\begin{tabular}{@{}c|c|c@{}}
\schele{Bkt}&\schele{Sex}&\schele{ZIP3}\\\hline
30--39&F&926\\
30--39&F&926\\
30--39&F&926\\
60--69&M&900
\end{tabular}
\caption{$R_2$: drop \schele{Med}.}
\label{fig:rel-2}
\end{subfigure}
\hfill
\begin{subfigure}[t]{0.23\linewidth}
\centering
\begin{tabular}{@{}c|c|c|c@{}}
\schele{Bkt}&\schele{Sex}&\schele{ZIP3}&\schele{Med}\\\hline
30--39&F&926&\schele{ART}\\
30--39&F&926&\schele{ART}\\
60--69&M&900&\schele{Ins.}
\end{tabular}
\caption{$R_3$: delete $t_3$.}
\label{fig:rel-3}
\end{subfigure}

\vspace{0.35ex}
{\footnotesize Source relation and three prepared releases.}
\end{minipage}

\vspace{0.25ex}

\begin{subfigure}[t]{\linewidth}
\centering
\begin{tikzpicture}[
    x=1cm,y=1cm,
    every node/.style={align=center},
    box/.style={
        draw,
        rounded corners=2pt,
        inner sep=3pt,
        minimum height=0.62cm,
        font=\scriptsize
    },
    sbox/.style={
        draw,
        rounded corners=2pt,
        inner sep=3pt,
        minimum height=0.98cm,
        text width=3.05cm,
        font=\scriptsize
    },
    prior/.style={box, fill=softblue, text width=2.75cm},
    rel/.style={box, fill=white, minimum width=2.35cm},
    repchg/.style={box, fill=softgray, minimum width=2.35cm},
    delchg/.style={box, fill=softred, minimum width=2.35cm},
    compat/.style={sbox, fill=softblue},
    good/.style={sbox, fill=softgreen},
    warn/.style={sbox, fill=softred},
    note/.style={font=\scriptsize, text=black!70},
    arr/.style={-{Latex[length=2mm]}, thick}
]

\node[prior] (target) at (0,1.15)
{target $t_1$\\ secret $s_{\mathsf{HIV}}$};

\node[prior] (profile) at (0,0.05)
{prior profile $Q$\\ $(30\text{--}39,F,926)$};

\node[prior] (group) at (0,-1.05)
{$R_Q(t_1)=\{t_1,t_2,t_3\}$\\ rate $=2/3$};

\draw[arr] (target) -- (profile);
\draw[arr] (profile) -- (group);

\node[note, font=\scriptsize\bfseries\color{black}] at (0,1.85) {Common input};

\node[note, font=\scriptsize\bfseries\color{black}] at (9.0,1.85) {Release $\rightarrow$ compatibility $\rightarrow$ disclosure};

\node[rel]    (r1) at (5.0,1.15) {$R_1$\\ keep \schele{Med}};
\node[repchg] (r2) at (9.0,1.15) {$R_2$\\ drop \schele{Med}};
\node[delchg] (r3) at (13.0,1.15) {$R_3$\\ delete $t_3$};

\node[note] (shared) at (9.0,0.43)
{computed relative to $R_Q(t_1)=\{t_1,t_2,t_3\}$};

\node[compat] (c1) at (5.0,-0.45)
{$C_1(t_1)=\{t_1,t_2\}$\\
$\rho_s(C_1)=1$\\
$\epsilon_1=\log(3/2)$};

\node[good] (c2) at (9.0,-0.45)
{$C_2(t_1)=\{t_1,t_2,t_3\}$\\
$\rho_s(C_2)=2/3$\\
$\epsilon_2=0$\\
\textit{adds ambiguity}};

\node[warn] (c3) at (13.0,-0.45)
{$C_3(t_1)=\{t_1,t_2\}$\\
$\rho_s(C_3)=1$\\
$\epsilon_3=\log(3/2)$\\
\textit{removes ambiguity}};

\draw[arr] (r1) -- (c1);
\draw[arr] (r2) -- (c2);
\draw[arr] (r3) -- (c3);

\draw[dashed, thick, draw=black!30] (shared.south west) -- (c1.north west);
\draw[dashed, thick, draw=black!30] (shared.south) -- (c2.north);
\draw[dashed, thick, draw=black!30] (shared.south east) -- (c3.north east);

\node[box, fill=softblue, text width=9.8cm] (takeaway) at (9.0,-1.85)
{$R_2$ expands the comparison class, so disclosure returns to the prior.
$R_3$ deletes the only non-\schele{ART} tuple, so disclosure stays as in $R_1$.};

\end{tikzpicture}
\caption{Model interpretation for target $t_1$ and secret $s_{\mathsf{HIV}}$.}
\label{fig:running-semantics}
\end{subfigure}

\caption{Running example. (a)--(d)Curation steps.~(e) The prior profile identifies the comparison group $R_Q(t_1)$; each
release induces a compatibility set $C_i(t_1)$; disclosure compares
the secret rate in $C_i(t_1)$ against the prior rate in $R_Q(t_1)$.
Gray denotes representation change, crimson tuple deletion, and green
the privacy-improving outcome.}
\label{fig:running-combined}
\vspace{-1mm}
\end{figure*}

%% file: content/inference.tex
\section{Curation Plans and Disclosure Semantics}
\label{sec:model}

A curation workflow transforms a source relation into a release by
applying ordinary preparation operators. Let $R$ be the source
relation and let $p = o_\ell \circ \cdots \circ o_1$
be a curation plan. Each $o_i$ is a deterministic operator, such as
projecting attributes, generalizing values, selecting a population,
or deleting tuples. The prefix $p_i=o_i\circ\cdots\circ o_1$ produces
the intermediate release $R_i=p_i(R)$, with $R_0=R$.

The prefix view is essential. A curator constructs the release one
operator at a time, and each prefix may be inspected, committed,
shared, cached, or reused as part of a later pipeline. The privacy
question is therefore not only what the final release reveals. It is
also how much inferential support each prefix leaves for protected
facts.

\paragraph{Observation and reasoning.}
A privacy-aware curation framework has three components. The
curation component describes how a prefix $p_i$ transforms $R$ into
$R_i$. The observation component describes what the adversary sees.
In the default regime, the adversary observes the released relation
$R_i$ and has prior knowledge about a target, but does not observe
the internal operator sequence. In a plan-aware regime, the adversary
also observes a description of the curation process, such as a
transformation script, exclusion criteria, metadata, or intermediate
artifacts. This regime matters when plans are published for
reproducibility, audit, or regulatory transparency.

The reasoning component maps the adversary's observation to
disclosure about protected facts. In this paper, we instantiate this
component using compatibility sets. This is a vehicle for making the
operator-level problem concrete, not the only possible semantics.
A dependency-explicit semantics could reason under a declared model
$\Sigma$ of dependencies or probabilistic rules; a Bayesian semantics
could use richer priors; and a multi-release semantics could account
for evidence accumulated across correlated outputs. The common
requirement is that the semantics explain how each curation step
changes the evidence available to the observer.

\paragraph{Secrets and prior visibility.}
A protected secret is a Boolean predicate over source tuples. For a
tuple $t$, $s(t)=1$ means that the protected fact holds for $t$. In
Example~\ref{ex:running}, the secret
\[
    s_{\mathsf{HIV}}(t)=1
    \quad\Longleftrightarrow\quad
    t[\schele{Diagnosis}]=\schele{HIV}
\]
captures whether the target has diagnosis \schele{HIV}. Other
secrets may describe membership in a cohort, the presence of a
treatment class, or any application-specific property of a tuple.

The observer also has prior information about the target. We model
this by a set $Q$ of visible attributes. The observer knows $t[Q]$
before seeing the release. The corresponding prior group is
\[
    R_Q(t)=\{u\in R : u[Q]=t[Q]\}.
\]
For E.x.~\ref{ex:running}, take
$Q=\{\schele{Bkt},\schele{Sex},\schele{ZIP3}\}$. For target $t_1$,
the prior group consists of $t_1,t_2,t_3$: the women in their
thirties in ZIP prefix 926.

\paragraph{Released representations.}
A curation prefix changes how a source tuple appears in the release.
We write $\pi_i(u)$ for the representation of source tuple $u$ after
prefix $p_i$. Projection removes fields from this representation.
Generalization replaces values by coarser values. Row operators do
not change the representation itself; they change whether the tuple
is present in the released relation.

Let $S_i\subseteq R$ be the source tuples that survive in $R_i$.
The release can then be viewed as the multiset
\[
    \{\pi_i(u):u\in S_i\}.
\]
This separates two effects that are often conflated:
representation-changing operators modify $\pi_i$, while
presence-changing operators modify $S_i$.

\paragraph{Compatibility-set instantiation.}
Under the compatibility-set semantics, the adversary compares the
target with source tuples that agree with the prior profile and have
the same released representation. The compatibility set after prefix
$p_i$ is
\[
    C_i(t)=
    \{u\in S_i :
      u[Q]=t[Q]\ \text{and}\ \pi_i(u)=\pi_i(t)\}.
\]
This is the set of source tuples still plausible as the target under
the observer's prior information and the visible release.

Fig.~\ref{fig:running-combined} gives the intuition. The top panel
shows the source and the three releases from Example~\ref{ex:running};
the bottom panel shows how the compatibility-set semantics reads the
same releases for target $t_1$. Keeping \schele{Med} visible in
$R_1$ compares the target only with the two ART tuples. Dropping
\schele{Med} in $R_2$ restores the larger comparison group and adds
ambiguity. Deleting $t_3$ in $R_3$ does the opposite: it removes the
only non-\schele{ART} tuple in the target profile, so the release is
smaller but not less revealing for this target.

\paragraph{Evidential disclosure.}
The compatibility set induces an empirical belief update. For
$X\subseteq R$, define
\[
    \rho_s(X)=
    \frac{|X\cap s^{-1}(1)|}{|X|}.
\]
The prior support for the secret is $\rho_s(R_Q(t))$. The support
after observing prefix $p_i$ is $\rho_s(C_i(t))$. We define
evidential disclosure as
\[
    \epsilon_i(s,t)
    =
    \log
    \frac{\rho_s(C_i(t))}
         {\rho_s(R_Q(t))}.
\]
A positive value means that the prefix increases support for the
secret relative to the prior profile. A value of zero means that the
release adds no support beyond the prior. A negative value means
that the release weakens support for the secret.

In the running example, the prior group for $t_1$ has HIV rate
$2/3$. Release $R_1$ raises the compatible secret rate to $1$, so
$\epsilon_1(s_{\mathsf{HIV}},t_1)=\log(3/2)$. Release $R_2$ returns
the comparison class to the prior group, so disclosure returns to
$0$. Release $R_3$ again leaves only the ART tuples compatible with
$t_1$, so disclosure remains the same as under $R_1$. The key point
is not the arithmetic itself, but the direction of evidence: $R_2$
adds ambiguity, while $R_3$ removes it.

\paragraph{Dependencies as evidence.}
The compatibility-set instantiation does not require the curator to
declare a dependency set. Dependencies enter through the secret rate
inside the target's comparison class. If \schele{Med}=\schele{ART}
is strongly associated with \schele{Diagnosis}=\schele{HIV}, then a
release that leaves the target compatible only with ART records
increases support for the HIV secret. Thus, the basic model treats
dependencies as evidential regularities in the source relation. A
dependency-explicit version can replace the empirical rate
$\rho_s(C_i(t))$ by inference under a declared semantics model
$\Sigma$.

\paragraph{Plan-level disclosure.}
The marginal effect of operator $o_i$ is
\[
    \delta_i(s,t)=\epsilon_i(s,t)-\epsilon_{i-1}(s,t).
\]
This quantity tells us whether the next curation step removed
evidence, added evidence, or left the inferential support unchanged.

\smallskip\noindent\textbf{Problem definition.} A release is evaluated over a set $\mathcal S$ of protected secrets
and a set $T$ of target tuples. We write
\[
    \alpha_i =
    \max_{s\in\mathcal S,\ t\in T} \epsilon_i(s,t)
\]
for the disclosure of prefix $p_i$. A curation plan satisfies budget
$\tau$ if every prefix remains below the budget:
\[
    \alpha_i \leq \tau
    \qquad\text{for all } i\in\{1,\ldots,\ell\}.
\]

The curation problem is to find a useful release without exceeding
this disclosure budget:
\[
    \max_{p\in\mathcal P} U(R_p)
    \quad
    \text{s.t.}
    \quad
    \alpha_i \leq \tau
    \text{ for every prefix } p_i\preceq p .
\]
Here $U$ is a task-specific utility measure, such as retained
attribute value, workload accuracy, or downstream model quality. The
prefix constraint captures the interactive setting: the curator needs
to know what the next operator does before committing to it. The
next section analyzes this marginal effect under the
compatibility-set instantiation.

%% file: content/leakagesurface.tex
\section{Operator Effects and Curation Guidance}
\label{sec:operators-guidance}

Under the compatibility-set instantiation, the privacy effect of a
curation operator becomes local. At prefix $p_i$, disclosure for a
target $t$ is determined by two objects: the representation map
$\pi_i$ and the surviving tuple set $S_i$. A candidate operator can
change one or both. Projection, generalization, and refinement act
primarily on $\pi_i$: they change how source tuples appear in the
release. Selection and deletion act primarily on $S_i$: they change
which tuples remain present. This distinction gives one concrete
explanation for why privacy under curation is non-monotone.

Fig.~\ref{fig:running-combined} shows this distinction. Dropping \schele{Med} in $R_2$ changes the representation of
the target profile: tuples that were separated by medication now
become comparable again, so the compatibility set for $t_1$ expands.
Deleting $t_3$ in $R_3$ changes presence instead: it removes the only
non-\schele{ART} tuple in the target profile. The two operations are
both ordinary curation steps, and both remove information in an
informal sense, but they have different effects on evidence. One
adds ambiguity; the other removes it.

\input{figs/planning}

Fig.~\ref{fig:interactive-planning} abstracts this example into the
interactive setting. At a current prefix, the curator has a set
$\Omega_i$ of candidate actions. The missing primitive is not merely
a final privacy test. It is a way to score each candidate before it
is committed: which compatibility sets can it change, in which
direction does it move disclosure, and what utility does it retain?

\paragraph{Representation-changing operators.}
If an operator changes $\pi_i$, it changes the comparison used by the
observer to relate the target to source tuples. A projection or
generalization merges representation classes. For a target $t$, the
new compatible set has the form
\[
    C_{i+1}(t)=C_i(t)\cup A,
\]
where $A$ is the set of newly indistinguishable tuples inside the
same visible profile. The new secret rate is
\[
    \rho_s(C_{i+1}(t))
    =
    \frac{|C_i(t)|\rho_s(C_i(t)) + |A|\rho_s(A)}
         {|C_i(t)|+|A|}.
\]
The direction of the privacy effect is therefore determined by the
tuples that join the target's comparison class. If the added tuples
have lower secret rate than $C_i(t)$, the operator weakens evidence
for the secret. If they have higher secret rate, it strengthens it.
Refinement has the reverse form: it splits a class and may remove
the tuples that were providing ambiguity.

This is the case for $R_2$ in Fig.~\ref{fig:running-combined}.
Dropping \schele{Med} merges the \schele{ART} and
\schele{None} representations inside the target's visible profile.
The compatible set for $t_1$ expands from $\{t_1,t_2\}$ to
$\{t_1,t_2,t_3\}$. Since the added tuple $t_3$ does not satisfy
$s_{\mathsf{HIV}}$, the secret rate falls from $1$ to $2/3$, and the
disclosure returns to the prior level. The operator helps because it
adds ambiguity.

\paragraph{Presence-changing operators.}
Selection and deletion behave differently. They change $S_i$, not
the released representation itself. If a removed tuple $d$ is not in
$C_i(t)$, then the target's evidence does not change:
\[
    C_{i+1}(t)=C_i(t).
\]
If $d\in C_i(t)$, then
\[
    C_{i+1}(t)=C_i(t)\setminus\{d\}.
\]
The sign of the privacy effect depends on what $d$ contributed. If
$s(d)=1$, the operator removes supporting evidence and may reduce
disclosure. If $s(d)=0$, it removes contrary evidence and may
increase disclosure. Equivalently, for $C=C_i(t)$,
\[
    \rho_s(C\setminus\{d\})
    =
    \frac{|C|\rho_s(C)-s(d)}
         {|C|-1}.
\]
Tuple deletion is therefore not inherently privacy-improving. It is
protective when it removes evidence for the secret, and harmful when
it removes the ambiguity that shielded the target.

This is the case for $R_3$ in Fig.~\ref{fig:running-combined}.
Tuple $t_3$ is the only matching non-\schele{ART} tuple in the
target's visible profile. Deleting it does not remove the visible ART
evidence; it removes the tuple that made the profile mixed. The
release becomes smaller, but the evidence for \schele{HIV} remains
concentrated on the two ART records. Hence $R_3$ has the same
disclosure for $t_1$ as $R_1$.

\paragraph{Local guidance.}
The same local view gives the primitive needed for interactive
curation. For each candidate next operator $o\in\Omega_i$, the
framework should identify the targets whose compatibility sets can
change. Representation-changing operators affect targets whose
classes are merged or split. Presence-changing operators affect only
targets whose compatibility sets contain removed tuples. All other
targets have zero marginal change under the compatibility-set
semantics.

For affected targets, the candidate induces a marginal disclosure
\[
    \delta_{i+1}^{o}(s,t)
    =
    \epsilon_{i+1}^{o}(s,t)-\epsilon_i(s,t),
\]
and a candidate prefix disclosure
\[
    \alpha_{i+1}^{o}
    =
    \max_{s\in\mathcal S,\ t\in T}
    \epsilon_{i+1}^{o}(s,t).
\]
The candidate is feasible only if
$\alpha_{i+1}^{o}\leq\tau$. Among feasible candidates, the curator
can compare utility gains. This gives the local primitive shown in
Fig.~\ref{fig:interactive-planning}: score candidate actions by
privacy and utility, keep only those below the disclosure budget, and
commit one action to obtain the next prefix.

Beyond rejecting unsafe final releases, the curator
should see what each proposed operator does before committing to it.
In the running example, the guidance loop distinguishes the two
one-step refinements of $R_1$: dropping \schele{Med} changes the
representation and adds ambiguity, while deleting $t_3$ changes
presence and removes ambiguity. The former moves the prefix toward
the disclosure budget; the latter does no privacy work for the
target.

This analysis is specific to the compatibility-set instantiation, but
the requirement is broader. Any disclosure semantics for
privacy-aware curation should support the same operator question:
given the current prefix and a candidate next step, what evidence
does the step add, remove, or leave unchanged?

%% file: figs/planning.tex
\begin{figure*}[t]
\centering
\footnotesize
\begin{tikzpicture}[
    x=1cm,y=1cm,
    every node/.style={align=center},
    main/.style={
        draw,
        rounded corners=3pt,
        fill=softblue,
        inner sep=5pt,
        minimum height=0.92cm,
        text width=2.35cm,
        font=\scriptsize
    },
    action/.style={
        draw,
        rounded corners=2pt,
        inner sep=3pt,
        minimum height=0.68cm,
        text width=1.95cm,
        font=\scriptsize
    },
    rep/.style={action, fill=softgray},
    val/.style={action, fill=softyellow},
    row/.style={action, fill=softred},
    extact/.style={action, fill=white, dashed, draw=black!55},
    score/.style={
        draw,
        rounded corners=2pt,
        inner sep=4pt,
        minimum height=0.66cm,
        text width=2.35cm,
        font=\scriptsize
    },
    panel/.style={
        draw=black!55,
        rounded corners=4pt,
        inner sep=6pt
    },
    title/.style={font=\scriptsize\bfseries, text=black},
    obj/.style={
        draw,
        rounded corners=3pt,
        fill=softblue,
        inner sep=5pt,
        text width=10.2cm,
        font=\scriptsize
    },
    future/.style={
        draw,
        dashed,
        rounded corners=3pt,
        fill=white,
        inner sep=4pt,
        text width=2.75cm,
        font=\scriptsize
    },
    badge/.style={
        circle,
        draw=black!70,
        fill=white,
        inner sep=1.2pt,
        font=\scriptsize\bfseries
    },
    arr/.style={-{Latex[length=2mm]}, thick},
    darr/.style={-{Latex[length=2mm]}, thick, dashed, draw=black!55}
]


\node[main] (prefix) at (-1.5,-0.2)
{\textbf{current prefix}\\[-0.2mm]
$p_i$ on $R$\\
$(\pi_i,S_i,\alpha_i)$};

\node[rep] (drop) at (2.00,0.50)
{\textbf{Project/drop}\\
drop attributes\\
$\pi_i$ changes};

\node[val] (gen) at (4.25,0.50)
{\textbf{Generalize}\\
coarsen values\\
$\pi_i$ changes};

\node[row] (sel) at (2.00,-0.65)
{\textbf{Select/delete}\\
filter/suppress tuples\\
$S_i$ changes};

\node[extact] (ins) at (4.25,-0.65)
{\textbf{Insert/impute}\\
add/synthesize values\\
extension};

\begin{scope}[on background layer]
  \node[
      panel,
      fit=(drop)(gen)(sel)(ins),
      label={[title]above:candidate curator actions $\Omega_i$}
  ] (actions) {};
\end{scope}

\node[score, fill=softgreen] (priv) at (7.75,0.70)
{\textbf{privacy}\\
affected $C_i(t)$\\
$\delta_{i+1}^{o}$};

\node[score, fill=softgreen] (util) at (7.75,-0.35)
{\textbf{utility}\\
candidate gain\\
$\Delta U(o)$};

\node[score, fill=white] (filter) at (7.75,-1.30)
{feasible if\\
$\alpha_{i+1}^{o}\leq\tau$};

\begin{scope}[on background layer]
  \node[
      panel,
      fit=(priv)(util)(filter),
      label={[title]above:score candidates}
  ] (scoring) {};
\end{scope}

\node[main, fill=softgreen, text width=2.55cm] (choose) at (11.75,1)
{\textbf{choose} $o^\star$\\
best feasible\\
utility--privacy score};

\node[main, text width=2.55cm] (next) at (11.75,-0.5)
{\textbf{next prefix}\\
$p_{i+1}=o^\star\circ p_i$\\
becomes new state};

\begin{scope}[on background layer]
  \node[
      panel,
      fit=(choose)(next),
      label={[title]above:local step}
  ] (update) {};
\end{scope}

\draw[arr] (prefix.east) -- (actions.west);
\draw[arr] (actions.east) -- (scoring.west);
\draw[arr] (scoring.east) -- (update.west);
\draw[arr] (choose) -- (next);


\node[obj] (objective) at (2.65,-2.60)
{\textbf{Problem:}
maximize $U(R_p)$ subject to prefix budget
$\alpha_j\leq\tau$ for every prefix $p_j\preceq p$.
Exact global search is combinatorial; local candidate scoring is the
interactive primitive.};

\node[future] (drift) at (11.75,-2.50)
{re-execution\\
under drift\\
check $p$ on evolved $R'$};

\node[future] (extmodel) at (11.75,-4.00)
{beyond the basic\\
compatibility model\\
multiple releases, $\Sigma$,\\
richer adversaries};

\begin{scope}[on background layer]
  \node[
      panel,
      fit=(drift)(extmodel),
      label={[title]above:extensions}
  ] (futurepanel) {};
\end{scope}

\draw[darr] (next.south) -- (drift.north);
\draw[darr] (filter.south east) |- (extmodel.west);

\node[badge] at ($(actions.north west)+(0.18,-0.18)$) {C1};
\node[badge] at ($(scoring.north west)+(0.18,-0.18)$) {C2};
\node[badge] at ($(objective.north west)+(0.18,-0.18)$) {C3};
\node[badge] at ($(drift.north west)+(0.18,-0.18)$) {C4};
\node[badge] at ($(extmodel.north west)+(0.18,-0.18)$) {C5};

\end{tikzpicture}

\caption{Interactive privacy-aware curation and its associated research
challenges.
}
\label{fig:interactive-planning}
\vspace{-1mm}
\end{figure*}
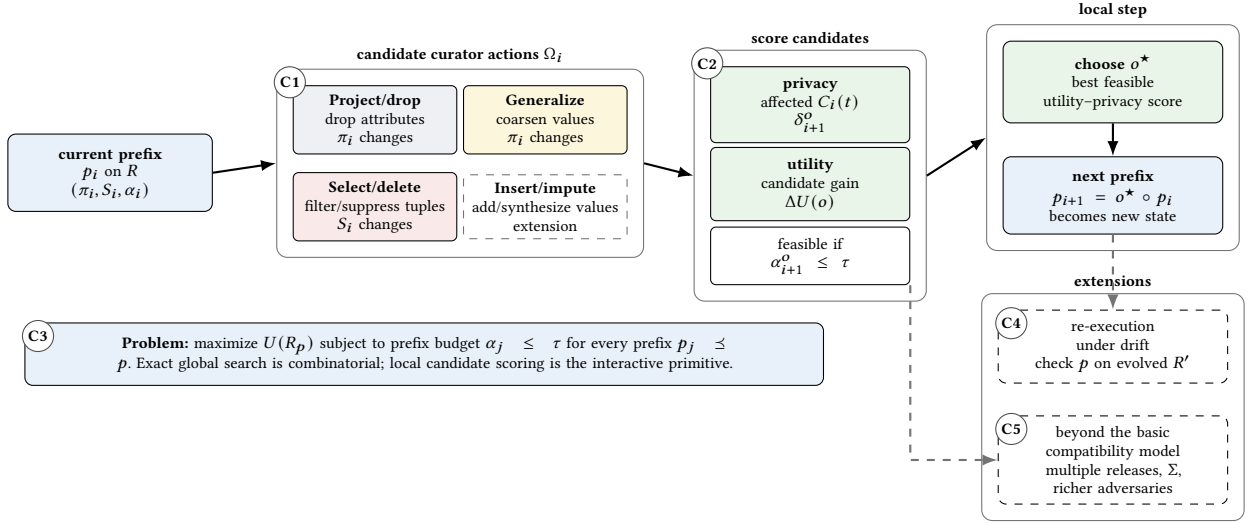

%% file: content/taxonomy.tex
\section{Research Challenges}
\label{sec:challenges}

The compatibility-set view gives privacy-aware curation a concrete
semantic object, but it also exposes why the problem is not a direct
application of existing release-level privacy tests. A curation plan
is not only a final table. It is a sequence of choices, and each
choice changes the evidence available to an observer. Fig.~\ref{fig:interactive-planning}
marks the four challenges that follow.

\paragraph{\textbf{C1. Per-operator privacy semantics.}}
The first challenge is to give each curation operator a privacy
semantics. Existing preparation systems already treat curation as a
sequence of typed transformations, but the types are usually
understood operationally: project removes attributes, generalize
coarsens values, select restricts the population, and delete removes
tuples. For privacy, these descriptions are too coarse. In
Example~\ref{ex:running}, both $R_2$ and $R_3$ are one-step
refinements of $R_1$, and both can be described informally as
``removing information.'' Yet they have different privacy effects.
Dropping \schele{Med} in $R_2$ changes the representation of the
target profile and merges the \schele{ART} and \schele{None} records,
thereby adding ambiguity. Deleting $t_3$ in $R_3$ changes tuple
presence and removes the only non-\schele{ART} record, thereby
removing ambiguity. Thus, the operator name alone does not determine
the privacy effect.

What is needed is a semantics that predicts, before an operator is
committed, whether it expands, shrinks, merges, or splits the
evidence classes that determine disclosure. In the compatibility-set
instantiation, this is the role of the representation/presence
distinction from Sec.~\ref{sec:operators-guidance}: column and value
operators act on how tuples are represented, while row operators act
on which tuples remain available as evidence.

\paragraph{\textbf{C2. Prefix-level accounting.}}
The second challenge is that disclosure must be accounted for over
plan prefixes, not only at the endpoint. Example~\ref{ex:running}
already exhibits the phenomenon. Suppose the curator eventually
publishes $R_2$, where \schele{Med} has been dropped. For target
$t_1$, the final disclosure is $0$: the compatible set is again
$\{t_1,t_2,t_3\}$, and the secret rate returns to the prior. But one
natural construction path reaches $R_2$ through $R_1$: first drop
\schele{Diag}, bucketize \schele{Age}, and shorten \schele{ZIP}; only
later drop \schele{Med}. Along this path, the intermediate prefix
$R_1$ has disclosure $\log(3/2)$, because the target is compatible
only with the two \schele{ART} tuples. Thus, the final release may be
acceptable even though the curation process passed through a more
revealing state.

This matters in interactive settings, where partial results may be
inspected, cached, shared with collaborators, or used to decide the
next transformation. It also matters conceptually: if the privacy
effect of an operator is its marginal change
$\delta_i(s,t)=\epsilon_i(s,t)-\epsilon_{i-1}(s,t)$, then the process
cannot be reconstructed from the endpoint alone. A privacy-aware
curation system should therefore maintain a trajectory
$\alpha_0,\alpha_1,\ldots,\alpha_\ell$, not only a final value
$\alpha_\ell$.

\paragraph{\textbf{C3. Guidance under a disclosure budget.}}
The third challenge is to make the semantics useful to the curator.
A data scientist does not only ask whether a release is safe; she
asks which operator to apply next. This turns curation into a
budgeted search problem:
\[
    \max_{p\in\mathcal P} U(R_p)
    \quad
    \text{s.t.}
    \quad
    \alpha_i \leq \tau
    \text{ for every prefix } p_i\preceq p .
\]
Exact search is unlikely to be the right abstraction for interactive
curation. The plan space is combinatorial, and utility is usually
task-specific. The useful primitive is local: given the current
prefix and a small set of candidate operators, estimate the marginal
disclosure of each candidate, rule out those that exceed the budget,
and explain the remaining privacy--utility tradeoff. In
Example~\ref{ex:running}, this is the difference between telling the
curator that dropping \schele{Med} removes a visible indicator and
telling her that deleting $t_3$ does no privacy work for the target.

\paragraph{\textbf{C4. Re-execution under drift.}}
The fourth challenge is that curated releases are often not
one-shot artifacts. The same preparation plan may be committed once
and re-executed whenever the source changes. A plan that satisfied a
disclosure budget at commit time may fail later because new tuples
enter a visible profile, existing tuples change their representation,
or correlations between visible attributes and secrets shift. This
makes privacy a maintained property of the curation pipeline, not a
one-time certificate. The relevant question is not only whether
$p(R)$ is safe today, but whether the same plan remains safe on
$R'$ before the next release is published.

\paragraph{\textbf{C5. Beyond compatibility sets.}}
The compatibility-set semantics measures how the release changes the
target's plausible comparison class. Several extensions are natural.
Insert and impute operators introduce synthetic or externally sourced
values that are not captured by pure suppression and coarsening.
Multiple releases require modeling how evidence accumulates across
correlated outputs. A dependency-explicit version would replace the
empirical compatibility score by inference under a declared semantics
model $\Sigma$, connecting privacy-aware curation to inference-aware
deletion under dependencies~\cite{chakraborty2025deletion,chakraborty2026deletion}. These extensions should refine, not
replace, the operator-level view: the central question remains how
each curation step changes the evidence that a released artifact
provides about protected facts.

Together, these challenges outline the problem space of
privacy-aware curation. The aim is not merely to certify a final
table, but to make privacy visible during the construction of that
table: which operator changed disclosure, in which direction, for
which targets, and under what budget.

\smallskip\noindent\textbf{Acknowledgements.} Chakraborty acknowledges the support of the Hasso-Plattner-Institut.